\documentclass[10pt]{article}
\usepackage{amssymb}

\newtheorem{Lemma}{Lemma}[section]
\newtheorem{Theorem}[Lemma]{Theorem}
\newtheorem{Proposition}[Lemma]{Proposition}
\newtheorem{Corollary}[Lemma]{Corollary}

\newenvironment{Proof}
{\begin{trivlist} \item[]{\bf Proof. }}%
{\hspace*{\fill}$\rule{.3\baselineskip}{.35\baselineskip}$\end{trivlist}}

\makeatletter
\@addtoreset{equation}{section}
\makeatother

\newcommand{\C}{\mathbb{C}}

\newcommand{\R}{\mathbb{R}}
\newcommand{\Z}{\mathbb{Z}}

\newfam\bifam
\font\tenbi=cmmib10 scaled \magstep1
\font\sevenbi=cmmib10 at 11pt
\font\fivebi=cmmib10 at 6pt
\textfont\bifam = \tenbi
\scriptfont\bifam= \sevenbi
\scriptscriptfont\bifam= \fivebi

\usepackage[dvips]{epsfig}
\usepackage{graphicx}

\begin{document}

\title{\bf Nonlinear Schr\"{o}dinger lattices \\ I: Stability of discrete solitons}

\author{D.E. Pelinovsky$^1$, P.G. Kevrekidis$^2$, and D.J. Frantzeskakis$^3$, \\
{\small $^{1}$ Department of Mathematics, McMaster
University, Hamilton, Ontario, Canada, L8S 4K1} \\
{\small $^{2}$ Department of Mathematics,
University of Massachusetts, Amherst, Massachusetts, 01003-4515, USA} \\
{\small $^{3}$ Department of Physics, University of Athens,
Panepistimiopolis, Zografos, Athens 15784, Greece}
}

\date{\today}
\maketitle

\begin{abstract}
We consider the discrete solitons bifurcating 
from the anti-continuum limit of
the discrete nonlinear Schr\"{o}dinger (NLS) lattice. The discrete
soliton in the anti-continuum limit represents an arbitrary finite
superposition of {\em in-phase} or {\em anti-phase} excited nodes,
separated by an arbitrary sequence of empty nodes. 
By using stability analysis, we prove that
the discrete solitons are all unstable near the anti-continuum
limit, except for the solitons, which consist of alternating 
anti-phase excited nodes. We
classify analytically and confirm numerically the number of unstable
eigenvalues associated with each family of the discrete solitons.
\end{abstract}

\section{Introduction}

Nonlinear instabilities and emergence of coherent structures in
differential-difference equations have become topics of physical
importance and mathematical interest in the past decade. Numerous
applications of these problems have emerged ranging from nonlinear
optics, in the dynamics of guided waves in inhomogeneous optical
structures \cite{EMSAPA02,PMAAESPL02} and photonic crystal lattices
\cite{ESCFS02,SKEA03}, to atomic physics, in the dynamics of
Bose-Einstein condensate droplets in periodic (optical lattice)
potentials \cite{CBFMMTSI01,CFFFMI03,ABDKS01,AKKS02} and from
condensed matter, in Josephson-junction ladders \cite{F03,MO03}, to
biophysics, in various models of the DNA double strand
\cite{DPB93,PDHW93}. This large range of models and applications has
been summarized by now in a variety of reviews such as
\cite{A97,FW98,HT99,KRB01,EJ03}.

One of the prototypical differential-difference models that is both
physically relevant and mathematically tractable is the so-called
discrete nonlinear Schr{\"o}dinger (NLS) equation,
\begin{equation}
\label{dNLS} i \dot{u}_n + \beta \Delta_d u_n + \gamma |u_n|^2 u_n =
0,
\end{equation}
where $u_n = u_n(t)$ is a complex amplitude in time $t$, $n \in
\Z^d$ is the $d$-dimensional lattice, $\Delta_d$ is the
$d$-dimensional discrete Laplacian, $\beta$ is the dispersion
coefficient, and $\gamma$ is the nonlinearity coefficient. Before we
delve into mathematical analysis of the discrete NLS equation, it
would be relevant to discuss briefly the recent physical
applications of this model.

The most direct implementation of the discrete NLS equation can be
identified in one-dimensional arrays of coupled optical waveguides
\cite{EMSAPA02,PMAAESPL02}. These may be multi-core structures
created in a slab of a semiconductor material (such as AlGaAs), or
virtual ones, induced by a set of laser beams illuminating a
photorefractive crystal. In this experimental implementation, there
are about forty lattice sites (guiding cores), and the localized
modes (discrete solitons) may propagate over twenty diffraction
lengths.

Light-induced photonic lattices \cite{ESCFS02,SKEA03} have
recently emerged as another application of the discrete NLS equation. 
The refractive index of the nonlinear photonic lattices changes
periodically due to a grid of strong beams, while a weaker probe
beam is used to monitor the localized modes (discrete solitons). A
number of promising experimental studies of discrete solitons in
light-induced photonic lattices was reported recently in physics
literature.

An array of Bose-Einstein condensate droplets trapped in a strong
optical lattice with thousands of atoms in each droplet, is another
direct physical realization of the discrete NLS equation
\cite{CBFMMTSI01,CFFFMI03}. In this context, the model can be
derived systematically by using the Wannier function expansions
\cite{ABDKS01,AKKS02}.

Besides applications to optical waveguides, photonic crystal
lattices, and Bose-Einstein condensates trapped in optical lattices,
the discrete NLS equation also arises as the envelope wave reduction
of the general nonlinear Klein-Gordon lattices \cite{PK92}.

This rich variety of physical contexts makes it timely and relevant
to analyze the mathematical aspects of the discrete NLS equation
(\ref{dNLS}), including the existence and stability of localized
modes (discrete solitons). A very helpful tool for such an analysis
is the so-called anti-continuum limit $\beta \to 0$ \cite{A97},
where the nonlinear oscillators of the model are uncoupled.
Existence of localized modes in this limit can be characterized in
full details \cite{HT99}. Persistence, multiplicity, and stability of
localized modes can be studied with continuation methods both 
analytically and numerically \cite{ABK04}.

We study localized modes of the discrete NLS equation (\ref{dNLS})
in the forthcoming series of two papers. This first paper 
describes stability analysis of 
discrete solitons in the one-dimensional NLS lattice
($d = 1$). The second paper will present Lyapunov--Schmidt
reductions for persistence, multiplicity and stability of discrete
vortices in the two-dimensional NLS lattice ($d = 2$).

This paper is structured as follows. We review the known results on
existence of one-dimensional discrete solitons in Section 2. General
stability and instability results for discrete solitons in the
anti-continuum limit are proved in Section 3. These stability
results are illustrated for two particular families of the discrete
solitons in Sections 4 and 5. Besides explicit perturbation series
expansions results, we compare asymptotic approximations and numerical
computations of stable and unstable eigenvalues in the linearized stability
problem. Section 6 concludes the first paper of this series.

\section{Existence of discrete solitons}

We consider the normalized form of the discrete NLS equation
(\ref{dNLS}) in one dimension ($d = 1$):
\begin{equation}
\label{NLS} i \dot{u}_n + \epsilon \left( u_{n+1} - 2 u_n +
u_{n-1} \right) + |u_n|^2 u_n = 0,
\end{equation}
where $u_n(t) : \R_+ \to \C$, $n \in \Z$, and $\epsilon > 0$ is the
inverse squared step size of the discrete one-dimensional NLS
lattice. The discrete solitons are given by the time-periodic
solutions of the discrete NLS equation (\ref{NLS}):
\begin{equation}
\label{soliton-form}
u_n(t) = \phi_n e^{i (\mu - 2 \epsilon) t + i \theta_0}, \qquad
\mu \in \R, \quad \phi_n \in \C, \quad n \in \Z,
\end{equation}
where $\theta_0 \in \R$ is parameter and $(\mu,\phi_n)$ solve the
nonlinear difference equations on $n \in \Z$:
\begin{equation}
\label{difference} (\mu - |\phi_n|^2) \phi_n = \epsilon \left(
\phi_{n+1} + \phi_{n-1} \right).
\end{equation}
Existence of the discrete solitons was studied recently in
\cite{DM00,BBJ00,BBV02}, inspired by the pioneer papers
\cite{AA90,MA94}. Recent summary of the existence results is 
given in \cite{ABK04}. Since discrete solitons in the 
focusing NLS lattice (\ref{NLS}) may
exist only for $\mu > 0$ \cite{ABK04} and the parameter $\mu$ is
scaled out by the scaling transformation,
\begin{equation}
\label{scaling-transformation} \phi_n = \sqrt{\mu} \hat{\phi}_n,
\qquad \epsilon = \mu \hat{\epsilon},
\end{equation}
the parameter $\mu > 0$ will henceforth be set to $\mu = 1$. Another
arbitrary parameter $\theta_0$, which exists due to the gauge
invariance of the discrete NLS equation (\ref{NLS}), is incorporated
in the anzats (\ref{soliton-form}), such that at least one value of 
$\phi_n$ can be chosen real-valued without lack of generality. Using this
convention, we represent below the known existence results.

\begin{Proposition}
\label{proposition-existence} There exist $\epsilon_0 > 0$, $\kappa
> 0$ and $\phi_{\infty} > 0$, such that the difference equations
(\ref{difference}) with $\mu = 1$ and $0 < \epsilon < \epsilon_0$
have continuous families of the discrete solitons with the
properties:
\begin{eqnarray}
\label{soliton} & (i) & \quad \lim_{\epsilon \to 0^+} \phi_n =
\phi_n^{(0)} = \left\{ \begin{array}{cc} e^{i \theta_n}, \quad n \in S, \\
0, \quad n \in \Z \backslash S, \end{array} \right. \\
\label{decay} & (ii) & \quad \lim_{|n| \to \infty} e^{\kappa |n|}
|\phi_n| = \phi_{\infty}, \\
\label{reality} & (iii) & \quad \phi_n \in \R, \; n \in \Z,
\end{eqnarray}
where $S$ is a finite set of nodes of the lattice $n \in \Z$ and
$\theta_n = \{ 0, \pi \}$, $n \in S$.
\end{Proposition}

\begin{Proof}
See Theorem 2.1 and Appendices A and B in \cite{ABK04} for the proof
of the limiting solution (\ref{soliton}) from the inverse function
theorem. See Theorem 3 in \cite{MA94} for the proof of the
exponential decay (\ref{decay}) from the bound estimates. See
Section 3.2 in \cite{HT99} for the proof of the reality condition
(\ref{reality}) from the conservation of the density current.
Various theoretical and numerical bounds on $\epsilon_0$ are
obtained in \cite{HT99,ABK04,DM00,MA94}.
\end{Proof}

Due to the property (\ref{reality}), the difference equations
(\ref{difference}) with $\mu = 1$ can be rewritten as follows:
\begin{equation}
\label{1difference} (1 - \phi_n^2) \phi_n = \epsilon \left(
\phi_{n+1} + \phi_{n-1} \right), \qquad n \in \Z.
\end{equation}
For our analysis, we shall derive two technical results on
properties of solutions $\phi_n$, $n \in \Z$.

\begin{Lemma}
\label{lemma-Taylor} There exists $\epsilon_0 > 0$, such that the
solution $\phi_n$, $n \in Z$ is represented by the convergent 
power series for $0 \leq \epsilon < \epsilon_0$:
\begin{equation}
\label{perturbation1} \phi_n = \phi_n^{(0)} + \sum_{k = 1}^{\infty}
\epsilon^k \phi_n^{(k)},
\end{equation}
where $\phi_n^{(0)}$ is given by (\ref{soliton}).
\end{Lemma}

\begin{Proof}
The statement follows from the Implicit Function Theorem (see
Theorem 2.7.2 in \cite{N74}), since the Jacobian matrix for the 
system (\ref{1difference}) is non-singular at $\phi_n =
\phi_n^{(0)}$, $n \in \Z$, while the right-hand-side of
the system (\ref{1difference}) is analytic in $\epsilon$.
\end{Proof}

\begin{Lemma}
\label{lemma-nodes} There exists $0 < \epsilon_1 < \epsilon_0$, such
that the number of changes in the sign of $\phi_n$ on $n \in \Z$ for
$0 < \epsilon < \epsilon_1$ equals to the number of
$\pi$-differences of the adjacent $\theta_n$, $n \in S$ in the
limiting solution (\ref{soliton}).
\end{Lemma}

\begin{Proof}
Consider two adjacent excited nodes $n_1, n_2 \in S$, separated by
$N$ empty nodes, such that $n_2 - n_1 = 1 + N$ and $N \geq 1$. We
need to prove that the number of $\pi$-differences in the argument
of $\phi_n$, $n_1 \leq n \leq n_2$ for small $\epsilon > 0$ is
exactly one if $\theta_{n_2} - \theta_{n_1} = \pi$ and zero if
$\theta_{n_2} - \theta_{n_1} = 0$. To do so, we consider the
difference equations (\ref{1difference}) on $n_1 < n < n_2$ as the
$N$-by-$N$ matrix system:
\begin{equation}
{\cal A}_N \mbox{\boldmath $\phi$}_N = \epsilon {\bf b}_N,
\end{equation}
where $\mbox{\boldmath $\phi$}_N =
(\phi_{n_1+1},...,\phi_{n_2-1})^T$, and
\begin{equation}
\nonumber {\cal A}_N = \left( \begin{array}{ccccc} 1 -
\phi_{n_1+1}^2 & - \epsilon & 0 & ... & 0 \\ -\epsilon & 1 -
\phi_{n_1+2}^2 & - \epsilon & ... & 0 \\ \vdots & \vdots & \vdots &
... & \vdots \\ 0 & 0 & 0 & ... & 1 - \phi_{n_2-1}^2 \end{array}
\right), \qquad {\bf b}_N = \left( \begin{array}{cc} \phi_{n_1} \\ 0
\\ \vdots \\ \phi_{n_2} \end{array} \right).
\end{equation}
Let $D_{I,J}$, $1 \leq I \leq J \leq N$ be the determinant of the
block of the matrix ${\cal A}_N$ between $I$-th and $J$-th rows and
columns. By Cramer's rule, we have
\begin{equation}
\phi_{n_1 + j} = \frac{\epsilon^j \phi_{n_1} D_{j+1,N} +
\epsilon^{N-j+1} \phi_{n_2} D_{1,N-j}}{D_{1,N}}.
\end{equation}
Since $\lim_{\epsilon \to 0} D_{I,J} = 1$ for all $1 \leq I \leq J
\leq N$, we have
\begin{eqnarray*}
\lim_{\epsilon \to 0} \epsilon^{-j} \phi_{n_1+j} = \phi_{n_1}, \quad
1 \leq j < \frac{N+1}{2}, \\
\lim_{\epsilon \to 0} \epsilon^{-j} \phi_{n_1+j} = \phi_{n_1} + \phi_{n_2},
\quad j = \frac{N+1}{2}, \\
\lim_{\epsilon \to 0} \epsilon^{j-1-N} \phi_{n_1+j} = \phi_{n_2},
\quad \frac{N+1}{2} < j \leq N.
\end{eqnarray*}
The statement of Lemma follows from the signs of $\phi_n$, 
$n_1 \leq n \leq n_2$ for small $\epsilon > 0$.
\end{Proof}

By Proposition \ref{proposition-existence} and
Lemma \ref{lemma-nodes}, all families of the discrete solitons as
$\epsilon \to 0$ can be classified by a sequence of $\{0\}$,
$\{+\}$, and $\{-\}$ of the limiting solution (\ref{soliton})
on the finite set $S$ \cite{ABK04}. In particular, 
we consider two ordered sets $S$:
\begin{equation}
\label{family-1} S_1 = \{ 1,2,3,...,N \}
\end{equation}
and
\begin{equation}
\label{family-2} S_2 = \{ 1,3,5,...,2N - 1 \},
\end{equation}
where ${\rm dim}(S_1) ={\rm dim}(S_2) = N < \infty$. The set
$S_1$ includes the Page mode ($N = 2$: $\theta_1 = \theta_2 = 0$)
and the twisted mode ($N = 2$: $\theta_1 = 0$, $\theta_2 = \pi$).
The set $S_2$ includes the Page and twisted modes ($N = 2$),
separated by an empty node.

\section{Stability of discrete solitons}

The spectral stability of discrete solitons is studied with the
standard linearization:
\begin{equation}
\label{linearization} u_n(t) = e^{i (1 - 2 \epsilon) t + i \theta_0}
\left( \phi_n + a_n e^{\lambda t} + \bar{b}_n e^{\bar{\lambda} t}
\right), \qquad \lambda \in \C, \quad (a_n,b_n) \in \C^2, \quad n
\in \Z,
\end{equation}
where $(\lambda,a_n,b_n)$ solve the linear eigenvalue problem on
$n \in \Z$:
\begin{eqnarray}
\nonumber \left(1 - 2 \phi_n^2\right) a_n - \phi_n^2 b_n -
\epsilon \left( a_{n+1} + a_{n-1} \right) & = & i \lambda a_n, \\
\label{a-b-problem} - \phi_n^2 a_n + \left(1 - 2 \phi_n^2\right) b_n
- \epsilon \left( b_{n+1} + b_{n-1} \right) & = & -i \lambda b_n.
\end{eqnarray}
The discrete soliton (\ref{soliton-form}) is called spectrally unstable
if there exists $\lambda$ and $(a_n,b_n)$, $n \in \Z$ in the problem
(\ref{a-b-problem}), such that ${\rm Re}(\lambda) > 0$ and $\sum_{n
\in \Z} \left( |a_n|^2 + |b_n|^2 \right) < \infty$. Otherwise, the
soliton is called weakly spectrally stable. Orbital stability of the
discrete one-pulse soliton was studied in the anti-continuum limit
$\epsilon \to 0$ \cite{W99} and close to the continuum limit
$\epsilon \to \infty$ \cite{KK01}. Spectral instabilities of
two-pulse and multi-pulse solitons were considered in
\cite{JA97,KBR01,KKM01,KW03,MJKA02,J04} by numerical and variational
approximations. It was well understood from
intuition supported by numerical simulations \cite{A97,MJKA02} 
that the discrete
solitons with the alternating sequence of $\theta_n = \{0,\pi\}$ in
the limiting solution (\ref{soliton}) are spectrally stable as
$\epsilon \to 0$ but have eigenvalues with so-called negative Krein 
signature, which become complex by means of the Hamiltonian--Hopf
bifurcations \cite{KBR01,J04}. All other families of discrete
solitons have unstable real eigenvalues $\lambda$ in the
anti-continuum limit for any $\epsilon \neq 0$ \cite{MJKA02}.

Here we prove these preliminary observations and find 
the precise number of stable and unstable eigenvalues 
in the linearized stability problem
(\ref{a-b-problem}) for small $\epsilon > 0$. 
Our results are similar to those in 
the Lyapunov-Schmidt reductions, which are applied to
continuous multi-pulse solitons in nonlinear Schr\"{o}dinger
equations \cite{SJA97,S98,K01,KK04}. In particular, the main
conclusion on stability of alternating up-down solitons and
instability of any other up-up and down-down sequences of solitons
was found for multi-pulse homoclinic orbits arising in the so-called
orbit-flip bifurcation \cite[p.176]{SJA97}. The same conclusion
agrees with qualitative predictions for the
discrete NLS equations \cite[p.66]{MJKA02}.

Let $\Omega = L^2(\Z,\C)$ be the Hilbert space of square-summable
bi-infinite complex-valued sequences $\{ u_n \}_{n \in \Z}$,
equipped with the inner product and norm:
\begin{equation}
\label{inner-product} ({\bf u},{\bf w})_{\Omega} = \sum_{n\in\Z}
\bar{u}_n w_n, \qquad \| {\bf u} \|^2_{\Omega} = \sum_{n \in \Z}
|u_n|^2 < \infty.
\end{equation}
We use bold notations ${\bf u}$ for an infinite-dimensional vector
in $\Omega$ that consists of components $u_n$ for all $n \in \Z$.
The stability problem (\ref{a-b-problem}) is transformed with the
substitution,
\begin{equation}
a_n = u_n + i w_n, \qquad b_n = u_n - i w_n, \qquad n \in \Z,
\end{equation}
to the form:
\begin{eqnarray}
\nonumber \left(1 - 3 \phi^2_n \right) u_n -
\epsilon \left( u_{n+1} + u_{n-1} \right) & = & - \lambda w_n, \\
\label{u-w-problem} \left(1 - \phi_n^2\right) w_n - \epsilon \left(
w_{n+1} + w_{n-1} \right) & = & \lambda u_n.
\end{eqnarray}
The matrix-vector form of the problem (\ref{u-w-problem}) is
\begin{equation}
\label{LL} {\cal L}_+ {\bf u} = - \lambda {\bf w}, \qquad {\cal L}_-
{\bf w} = \lambda {\bf u},
\end{equation}
where ${\cal L}_{\pm}$ are infinite-dimensional symmetric
tri-diagonal matrices, which consist of elements:
$$
\left( {\cal L}_+ \right)_{n,n} = 1 - 3 \phi_n^2, \qquad \left(
{\cal L}_- \right)_{n,n} = 1 - \phi_n^2, \qquad \left( {\cal
L}_{\pm} \right)_{n,n+1} = \left( {\cal L}_{\pm} \right)_{n+1,n} = -
\epsilon.
$$
Equivalently, the stability problem (\ref{LL}) is rewritten in the
Hamiltonian form:
\begin{equation}
\label{eigenvalue} {\cal J} {\cal H} \mbox{\boldmath $\psi$} =
\lambda \mbox{\boldmath $\psi$},
\end{equation}
where $\mbox{\boldmath $\psi$}$ is the infinite-dimensional
eigenvector, which consists of 2-blocks of $(u_n,w_n)^T$, ${\cal J}$
is the infinite-dimensional skew-symmetric matrix, which consists of
$2$-by-$2$ blocks of
$$
{\cal J}_{n,m} = \left( \begin{array}{cc} 0 & 1 \\ -1 & 0
\end{array} \right) \delta_{n,m},
$$
and ${\cal H}$ is the infinite-dimensional symmetric matrix, which
consists of $2$-by-$2$ blocks of
$$
{\cal H}_{n,m} = \left( \begin{array}{cc} ({\cal L}_+)_{n,m} & 0 \\
0 & ({\cal L}_-)_{n,m} \end{array} \right).
$$
The representation (\ref{eigenvalue}) follows from the Hamiltonian
structure of the discrete NLS equation (\ref{NLS}), where ${\cal J}$
is the symplectic operator and ${\cal H}$ is the linearized
Hamiltonian. By Lemma \ref{lemma-Taylor}, the matrix ${\cal H}$ is expanded into
the power series:
\begin{equation}
\label{matrix1} {\cal H} = {\cal H}^{(0)} + \sum_{k = 1}^{\infty}
\epsilon^k {\cal H}^{(k)},
\end{equation}
where ${\cal H}^{(0)}$ is diagonal with two blocks:
\begin{equation}
\label{energy-0}
{\cal H}_{n,n}^{(0)} = \left( \begin{array}{cc} -2 & 0 \\
0 & 0 \end{array} \right), \;\; n \in S, \qquad
{\cal H}_{n,n}^{(0)} = \left( \begin{array}{cc} 1 & 0 \\
0 & 1 \end{array} \right), \;\; n \in \Z \backslash S.
\end{equation}
Let $N = {\rm dim}(S) <\infty$. The spectrum of ${\cal H}^{(0)}
\mbox{\boldmath $\varphi$} = \gamma \mbox{\boldmath $\varphi$}$ has
exactly $N$ negative eigenvalues $\gamma = -2$, $N$ zero eigenvalues
$\gamma = 0$ and infinitely many positive eigenvalues $\gamma = +1$.
The negative and zero eigenvalues $\gamma = -2$ and $\gamma = 0$ map
to $N$ double zero eigenvalues $\lambda = 0$ in the eigenvalue
problem ${\cal J} {\cal H}^{(0)} \mbox{\boldmath $\psi$} = \lambda
\mbox{\boldmath $\psi$}$. The positive eigenvalues $\gamma = +1$ map
to the infinitely many eigenvalues $\lambda = \pm i$.

Since finitely many zero eigenvalues of
${\cal J} {\cal H}^{(0)}$ are isolated from the rest of the
spectrum, their shifts vanish as $\epsilon \to 0$, according to the
regular perturbation theory \cite{HJ85}. We can therefore locate
small unstable eigenvalues ${\rm Re}(\lambda) > 0$ of the stability
problem (\ref{eigenvalue}) for small $\epsilon > 0$ 
from their limits at $\epsilon = 0$. 
On the other hand, infinitely many imaginary eigenvalues of
${\cal J} {\cal H}^{(0)}$ become the continuous spectrum band as
$\epsilon \neq 0$ \cite{LL92}. However, since the difference
operator ${\cal J H}$ has exponentially decaying
potentials $\phi_n$, $n \in \Z$, due to the decay condition 
(\ref{decay}), the
continuous spectral bands of ${\cal J H}$ are located on the
imaginary axis of $\lambda$ near the points $\lambda = \pm i$,
similarly to the case $\phi_n = 0$, $n \in \Z$ \cite{LL92}.
Therefore, the infinite-dimensional part of the spectrum does not
produce any unstable eigenvalues ${\rm Re}(\lambda) > 0$ in the
stability problem (\ref{eigenvalue}) as $\epsilon > 0$. Results of 
the regular perturbation theory are formulated and proved below.

\begin{Lemma}
\label{proposition-stability} Assume that $\phi_n$, $n \in \Z$ is
the discrete soliton, described in Proposition
\ref{proposition-existence}. Let $N = {\rm dim}(S) < \infty$. Let
$\gamma_j$, $1 \leq j \leq N$ be small eigenvalues of ${\cal H}$ as
$\epsilon \to 0$, such that
\begin{equation}
\label{leading-order0} \lim_{\epsilon \to 0} \gamma_j = 0, \qquad
1 \leq j \leq N.
\end{equation}
There exists $0 < \epsilon_* \leq \epsilon_0$, such that the
eigenvalue problem (\ref{eigenvalue}) with $\phi_n$, $n \in \Z$ and
$0 < \epsilon < \epsilon_*$ has $N$ pairs of small eigenvalues
$\lambda_j$ and $-\lambda_j$, $1 \leq j \leq N$, 
that satisfy the leading-order behavior:
\begin{equation}
\label{leading-order} \lim_{\epsilon \to 0}
\frac{\lambda_j^2}{\gamma_j} = 2, \qquad 1 \leq j \leq N.
\end{equation}
\end{Lemma}

\begin{Proof}
Since the operator ${\cal L}_+$ is Fredholm of zero index and empty
kernel at $\epsilon = 0$, it can be inverted for small $\epsilon > 0$ 
and the non-self-adjoint eigenvalue
problem (\ref{LL}) can be transformed to the self-adjoint
diagonalization problem:
\begin{equation}
{\cal L}_- {\bf w} = - \lambda^2 {\cal L}^{-1}_+ {\bf w},
\end{equation}
such that 
\begin{equation}
\label{variational-principle} \lambda^2 = - \frac{({\bf w},{\cal
L}_- {\bf w})_{\Omega}}{({\bf w},{\cal L}^{-1}_+ {\bf w})_{\Omega}},
\end{equation}
where the inner product is defined in (\ref{inner-product}). 
Since all small eigenvalues of ${\cal H}$ are small eigenvalues
of ${\cal L}_-$, we denote ${\bf w}_j$ be an eigenvector of ${\cal L}_-$,
which corresponds to the small eigenvalue $\gamma_j$, $1 \leq j \leq
N$ in the limiting condition (\ref{leading-order0}). By continuity of the
eigenvectors and completeness of ${\rm ker}({\cal L}^{(0)}_-)$,
there exists a set of normalized coefficients $\{ c_{n,j} 
\}_{n \in S}$ for each $1 \leq j \leq N$, such that
\begin{equation}
\lim_{\epsilon \to 0} {\bf w}_j = {\bf w}_j^{(0)} = \sum_{n \in S}
c_{n,j} {\bf e}_n, \qquad \sum_{n \in S} |c_{n,j}|^2 = 1,
\end{equation}
where ${\bf e}_n$ is the unit vector in $\Omega$. It follows from
the direct computations that
\begin{equation}
\label{asymptotics} \lim_{\epsilon \to 0} ({\bf w}_j,{\cal L}^{-1}_+
{\bf w}_j) = ({\bf w}_j^{(0)},{\cal L}^{(0)-1}_+ {\bf w}_j^{(0)}) =
- \frac{1}{2}.
\end{equation}
The leading-order behavior (\ref{leading-order}) follows from
(\ref{variational-principle}) and (\ref{asymptotics}) by the regular
perturbation theory \cite{HJ85}.
\end{Proof}

\begin{Corollary}
\label{corollary-signature}
Each small positive eigenvalue $\gamma_j$ corresponds to 
a pair of positive and negative eigenvalues $\lambda_j$ and 
$-\lambda_j$ for small $\epsilon > 0$. Each small negative 
eigenvalue $\gamma_j$ corresponds to a pair of purely imaginary 
eigenvalues $\lambda_j$ and $-\lambda_j$ for small $\epsilon > 0$. 
The latter eigenvalues have negative Krein signature: 
\begin{equation}
\left( \mbox{\boldmath $\psi$}, {\cal H} \mbox{\boldmath $\psi$}
\right) = \left( {\bf u}, {\cal L}_+ {\bf u} \right) + 
\left( {\bf w}, {\cal L}_- {\bf w} \right) = 2 
\left( {\bf w}, {\cal L}_- {\bf w} \right) < 0.
\end{equation}
\end{Corollary}

For any $\epsilon \neq 0$, there exists a simple zero eigenvalue of
${\cal H}$ due to the gauge symmetry of the discrete solitons
(\ref{soliton-form}), as the parameter $\theta_0$ is arbitrary, 
such that ${\cal L}_- \mbox{\boldmath $\phi$} = {\bf 0}$. When
all other $(N-1)$ eigenvalues $\gamma_j$ are non-zero for any
$\epsilon \neq 0$, the splitting of the semi-simple zero eigenvalue
of ${\cal H}^{(0)}$ is called {\em generic}. The generic splitting
gives a sufficient condition for unique (up to the gauge invariance)
continuation of discrete solitons for $\epsilon \neq 0$
\cite{SJA97}, which is also garanteed by Proposition
\ref{proposition-existence} \cite{MA94}.

Let $n_0$ and $p_0$ be the numbers of negative and positive
eigenvalues $\gamma_j$, defined in Lemma
\ref{proposition-stability}. The splitting is generic if $p_0 = N -
1 - n_0$. The numbers $n_0$ and $p_0$ are computed exactly from the
limiting solution (\ref{soliton}) as follows.

\begin{Lemma}
\label{proposition-n0} There exists $0 < \epsilon_1 < \epsilon_0$,
such that the index $n_0$ for $0 < \epsilon < \epsilon_1$ equals to
the number of $\pi$-differences of the adjacent $\theta_n$, $n \in
S$ in the limiting solution (\ref{soliton}), while $p_0 = N - 1 -
n_0$.
\end{Lemma}

\begin{Proof}
Since ${\cal L}_- \mbox{\boldmath $\phi$} = {\bf 0}$ for 
any $0 < \epsilon < \epsilon_0$, the number $n_0$ of 
negative eigenvalues of ${\cal L}_-$ coincides with 
the number of times when $\mbox{\boldmath $\phi$}$ changes the
sign, by the Discrete Sturm--Liouville Theorem \cite{LL92}. 
In the case $\epsilon = 0$, this number equals the number of
$\pi$-differences of the adjacent $\theta_n$, $n \in S$ in the
limiting solution (\ref{soliton}). By Lemma \ref{lemma-nodes}, the
number remains continuous as $\epsilon \neq 0$. The difference
equation ${\cal L}_- {\bf w} = {\bf 0}$ has only two fundamental
solutions, such that ${\bf w} = c_1 {\bf w}_1 + c_2 {\bf w}_2$,
where $c_1, c_2$ are arbitrary parameters, ${\bf w}_1 =
\mbox{\boldmath $\phi$}$ is exponentially decaying as $|n| \to
\infty$, and ${\bf w}_2$ is exponentially growing as $|n| \to
\infty$, due to the discrete Wronskian identity \cite{LL92}. As
a result, the kernel of ${\cal L}_-$ is one-dimensional for
$\epsilon \neq 0$, such that $p_0 = N - 1 - n_0$.
\end{Proof}

It was recently studied \cite{P04,KKS04} that there exists a closure
relation between the negative index of the linearized Hamiltonian
${\cal H}$ and the number of unstable eigenvalues of the linearized
operator ${\cal J H}$. The closure relation can be extended from the
coupled NLS equations to the discrete NLS equations by using the
same methods \cite{P04,KKS04}. We hence formulate the closure
relation for the discrete NLS equations (\ref{NLS}).

\begin{Proposition}
\label{proposition-closure} Let $n({\cal H})$ be the finite number
of negative eigenvalues of ${\cal H}$. Let $N_{\rm real}$ be the
number of positive real eigenvalues $\lambda$ in the problem
(\ref{eigenvalue}), $N_{\rm imag}^-$ be the number of pairs of
purely imaginary eigenvalues $\lambda$ with negative Krein signature
$\left( \mbox{\boldmath $\psi$}, {\cal H} \mbox{\boldmath $\psi$}
\right) < 0$, and $N_{\rm comp}$ be the number of complex
eigenvalues $\lambda$ in the first open quadrant of $\lambda$. Let
$p(P') = 1$ if $P' \geq 0$ and $p(P') = 0$ if $P'<0$, where
\begin{equation}
\label{derivative-power} P' = \| \mbox{\boldmath $\phi$}
\|^2_{\Omega} - \epsilon \frac{d}{d \epsilon} \| \mbox{\boldmath
$\phi$} \|^2_{\Omega}.
\end{equation}
Assume that $\lambda = 0$ is a double eigenvalue of the problem
(\ref{eigenvalue}). Assume that no purely imaginary eigenvalues
$\lambda$ exist inside the continuous spectrum or have  zero
Krein signature. The indices above satisfy the closure relation:
\begin{equation}
\label{closure} n({\cal H}) - p(P') = N_{\rm real} + 2 N_{\rm
imag}^- + 2 N_{\rm comp}.
\end{equation}
\end{Proposition}

\begin{Proof}
The left-hand-side of (\ref{closure}) is the negative index of
${\cal H}$ in the constrained subspace of $\Omega$, which is reduced
by one, if the power $\| \mbox{\boldmath $\phi$} \|_{\Omega}^2$ is
increasing function of $\mu$. Due to the scaling transformation
(\ref{scaling-transformation}), the derivative of $\|
\mbox{\boldmath $\phi$} \|_{\Omega}^2$ in $\mu$ is given by
(\ref{derivative-power}), where the hats for $\phi_n$ and $\epsilon$
are omitted. The right-hand-side of (\ref{closure}) is the negative
index of ${\cal H}$ on the subspace of $\Omega$, associated to the
eigenvalue problem (\ref{LL}). The two indices are equal under the
assumptions of the proposition, according to \cite{P04,KKS04}.
\end{Proof}

\begin{Corollary}
\label{corollary-closure} There exists $0 < \epsilon_2 <
\epsilon_1$, such that the indices of Proposition
\ref{proposition-closure} for $0 < \epsilon < \epsilon_2$ equal to
the indices of Lemmas \ref{proposition-stability} and
\ref{proposition-n0} as follows:
\begin{equation}
\label{count} n({\cal H}) = N + n_0, \quad p(P') = 1, \quad N_{\rm
real} = N - 1 - n_0, \quad N_{\rm imag}^- = n_0, \quad N_{\rm comp}
= 0,
\end{equation}
and the closure relation (\ref{closure}) is met.
\end{Corollary}

When the assumptions of Proposition \ref{proposition-closure} are
not satisfied, instability bifurcations may occur in the eigenvalue
problem (\ref{eigenvalue}), which results in the redistribution of
the numbers $n({\cal H})$, $p(P')$, $N_{\rm real}$, $N_{\rm
imag}^-$, and $N_{\rm comp}$. The Hamiltonian--Hopf bifurcation,
which is typical for the discrete multi-humped solitons
\cite{KBR01,KW03,J04}, occurs when the purely imaginary eigenvalues
$\lambda$ of negative Krein signature $\left( \mbox{\boldmath
$\psi$}, {\cal H} \mbox{\boldmath $\psi$} \right) < 0$ collide with
the purely imaginary eigenvalues $\lambda$ of positive Krein
signature $\left( \mbox{\boldmath $\psi$}, {\cal H} \mbox{\boldmath
$\psi$} \right) > 0$ or with the continuous spectral band and
bifurcate as complex unstable eigenvalues $\lambda$ with ${\rm
Re}(\lambda) > 0$. It follows from Corollaries \ref{corollary-signature} 
and \ref{corollary-closure} that there can be 
at most $n_0$ Hamiltonian--Hopf instability
bifurcations, which result in at most $N + n_0 - 1$ unstable
eigenvalues, unless the indices $n({\cal H})$ and $p(P')$ change as
a result of the zero eigenvalue bifurcations.

Combining Lemmas \ref{proposition-stability} and
\ref{proposition-n0} and Corollaries \ref{corollary-signature} 
and \ref{corollary-closure}, we
summarize the main stability--instability result for the discrete
solitons of the discrete NLS equation (\ref{NLS}).

\begin{Theorem}
\label{theorem-stability} Let $n_0$ be the number of
$\pi$-differences of the adjacent $\theta_n$, $n \in S$ in the
limiting solution (\ref{soliton}). The discrete soliton is
spectrally stable for small $\epsilon > 0$ if
and only if $n_0 = N - 1$. When $n_0 < N -1$, the discrete soliton
is spectrally unstable with exactly $N-1-n_0$ real unstable
eigenvalues $\lambda$ in the problem (\ref{eigenvalue}). When $n_0
\neq 0$, there exists $n_0$ pairs of purely imaginary eigenvalues
$\lambda$ with negative Krein signature, which may bifurcate to
complex unstable eigenvalues $\lambda$ away from the anti-continuum
limit $\epsilon \to 0$.
\end{Theorem}

The splitting of the zero eigenvalue of ${\cal H}^{(0)}$, which
defines the stability-instability conclusion of Theorem
\ref{theorem-stability}, may occur in different powers of $\epsilon$
as $\epsilon \to 0$. The power of $\epsilon$, where it happens,
depends on the set $S$, which classifies the family of the discrete
solitons $\phi_n$, $n \in \Z$. For the sets $S_1$ and $S_2$, 
which are defined by (\ref{family-1}) and (\ref{family-2}), 
we show that the generic splitting of the zero eigenvalue 
occurs in the first and second orders of $\epsilon$, respectively. 
These results are reported in the next two sections.

\section{Bifurcations of the discrete solitons in the set $S_1$}

Here we study the set $S_1$ with the explicit 
perturbation series expansions. These methods illustrate the general
results of Theorem \ref{theorem-stability} and give 
asymptotic aproximations for stable and unstable eigenvalues 
of the linearized stability problem (\ref{a-b-problem}). 
We compare the asymptotic and numerical approximations 
in the simplest cases $N = 2$ and $N = 3$.

By Lemma \ref{lemma-Taylor}, solution of the difference
equations (\ref{1difference}) is defined by the power series
(\ref{perturbation1}), where $\phi_n^{(0)}$ is given by
(\ref{soliton}) with $\theta_n = \{0,\pi\}$ for all $n \in S$ and
$\phi_n^{(1)}$ solves the inhomogeneous problem:
\begin{equation}
\label{system1} (1 - 3 \phi_n^{(0) 2} ) \phi_n^{(1)} =
\phi_{n+1}^{(0)} + \phi_{n-1}^{(0)}, \qquad n \in \Z.
\end{equation}
For the set $S_1$, defined by (\ref{family-1}), the system
(\ref{system1}) has the unique solution:
\begin{eqnarray}
\nonumber \phi_n^{(1)} & = & - \frac{1}{2} \left( \cos
(\theta_{n-1}-\theta_n) + \cos(\theta_{n+1}-\theta_n) \right) e^{i
\theta_n}, \quad 2 \leq n \leq N-1, \\
\nonumber \phi_1^{(1)} & = & -\frac{1}{2} \cos(\theta_2 - \theta_1)
e^{i \theta_1}, \qquad \phi_N^{(1)} = -\frac{1}{2} \cos(\theta_N -
\theta_{N-1}) e^{i \theta_N}, \\
\label{solution1} \phi_0^{(1)} & = & e^{i \theta_1}, \qquad
\phi_{N+1}^{(1)} = e^{i \theta_N},
\end{eqnarray}
while all other elements of $\phi_n^{(1)}$ are empty. The symmetric
matrix ${\cal H}$ is defined by the power series 
(\ref{matrix1}), where ${\cal H}^{(0)}$ is given by (\ref{energy-0})
and ${\cal H}^{(1)}$ consists of blocks:
\begin{eqnarray}
{\cal H}_{n,n}^{(1)} = -2 \phi_n^{(0)} \phi_n^{(1)} \left(
\begin{array}{cc} 3 & 0 \\ 0 & 1 \end{array} \right), \qquad
\label{energy1} {\cal H}_{n,n+1}^{(1)} = {\cal
H}_{n+1,n}^{(1)} = -  \left( \begin{array}{cc} 1 & 0 \\
0 & 1 \end{array} \right).
\end{eqnarray}
while all other blocks of ${\cal H}_{n,m}^{(1)}$ are empty. The
semi-simple zero eigenvalue of the problem ${\cal
H} \mbox{\boldmath $\varphi$} = \gamma \mbox{\boldmath $\varphi$}$
is split as $\epsilon > 0$ according to the perturbation series
expansion:
\begin{eqnarray}
\label{eigenvector1} \mbox{\boldmath $\varphi$} = \mbox{\boldmath
$\varphi$}^{(0)} + \epsilon \mbox{\boldmath $\varphi$}^{(1)} + {\rm
O}(\epsilon^2), \qquad \gamma = \epsilon \gamma_1 + {\rm
O}(\epsilon^2).
\end{eqnarray}
Let $\gamma = 0$ be a semi-simple eigenvalue of ${\cal
H}^{(0)}$ with $N$ linearly independent eigenvectors 
${\bf f}_n$, $n \in S$. Recalling that $\sin \theta_n = 0$ 
and $\cos \theta_n = \pm 1$ for all $n \in S$, we 
normalize ${\bf f}_n$ by the only non-zero
block $(0,\cos \theta_n)^T$ at the $n$-th position, 
for the sake of convenience. The zero-order 
term $\mbox{\boldmath $\varphi$}^{(0)}$ takes the form:
\begin{equation}
\label{zeroeigenvector1} \mbox{\boldmath $\varphi$}^{(0)} = \sum_{n
\in S} c_n {\bf f}_n,
\end{equation}
where $c_n \in \C$, $n \in S$ are coefficients of the linear
superposition. The first-order term 
$\mbox{\boldmath $\varphi$}^{(1)}$ is found
from the inhomogeneous system:
\begin{equation}
\label{linearsystem1} {\cal H}^{(0)} \mbox{\boldmath
$\varphi$}^{(1)} = \gamma_1 \mbox{\boldmath $\varphi$}^{(0)} -
{\cal H}^{(1)} \mbox{\boldmath $\varphi$}^{(0)}.
\end{equation}
Projecting the system (\ref{linearsystem1}) onto the kernel of
${\cal H}^{(0)}$, we find that the first-order correction $\gamma_1$
is defined by the reduced eigenvalue problem:
\begin{equation}
\label{reducedeigenvalue1} {\cal M}_1 {\bf c} = \gamma_1 {\bf c},
\end{equation}
where ${\bf c} = (c_1,...,c_N)^T$ and ${\cal M}_1$ is a tri-diagonal
$N$-by-$N$ matrix, given by
\begin{equation}
\label{energyM1} \left( {\cal M}_1 \right)_{m,n} = \left( {\bf f}_m,
{\cal H}^{(1)} {\bf f}_n \right), \qquad 1 \leq n,m \leq N,
\end{equation}
or explicitly, based on the first-order solution (\ref{solution1}) and 
(\ref{energy1}):
\begin{eqnarray}
\nonumber (M_1)_{n,n} = \cos(\theta_{n+1} - \theta_n) +
\cos(\theta_{n-1} - \theta_n), \qquad 1 < n < N, \\
\nonumber
(M_1)_{n,n+1} = (M_1)_{n+1,n} = - \cos(\theta_{n+1}- \theta_n),
\qquad 1 \leq n < N, \\
\label{Melements1}
(M_1)_{1,1} = \cos(\theta_2 - \theta_1), \qquad (M_1)_{N,N} =
\cos(\theta_N - \theta_{N-1}).
\end{eqnarray}
Similarly, the multiple zero eigenvalue of the
problem ${\cal J H} \mbox{\boldmath $\psi$} = \lambda
\mbox{\boldmath $\psi$}$ is split as $\epsilon > 0$ according
to the perturbation series expansion:
\begin{eqnarray}
\label{Lyapunoveigenvector1} \mbox{\boldmath $\psi$} =
\mbox{\boldmath $\psi$}^{(0)} + \sqrt{\epsilon} \mbox{\boldmath
$\psi$}^{(1)} + \epsilon \mbox{\boldmath $\psi$}^{(2)} + {\rm
O}(\epsilon \sqrt{\epsilon}), \qquad \lambda = \sqrt{\epsilon}
\lambda_1 + \epsilon \lambda_2 + {\rm O}(\epsilon \sqrt{\epsilon}).
\end{eqnarray}
Let $\lambda = 0$ be a multiple eigenvalue of ${\cal J} {\cal
H}^{(0)}$ with $N$ linearly independent eigenvectors ${\bf f}_n$, $n
\in S$ and $N$ linearly independent generalized eigenvectors ${\bf
g}_n$, $n \in S$. The eigenector ${\bf g}_n$ has the only non-zero block
$(\cos\theta_n,0)^T$ at the $n$-th position. The zero-order term 
is given by (\ref{zeroeigenvector1}) as $\mbox{\boldmath
$\psi$}^{(0)} = \mbox{\boldmath $\varphi$}^{(0)}$, while 
the first-order term $\mbox{\boldmath $\psi$}^{(1)}$ is given by
\begin{equation}
\label{Lyapunovzeroeigenvector1} \mbox{\boldmath $\psi$}^{(1)} =
\frac{\lambda_1}{2}\sum_{n \in S} c_n {\bf g}_n.
\end{equation}
The second-order term
$\mbox{\boldmath $\psi$}^{(2)}$ is found from the inhomogeneous
system:
\begin{equation}
\label{Lyapunovlinearsystem1} {\cal J} {\cal H}^{(0)}
\mbox{\boldmath $\psi$}^{(2)} = \lambda_1 \mbox{\boldmath
$\psi$}^{(1)} + \lambda_2 \mbox{\boldmath $\psi$}^{(0)} - {\cal J}
{\cal H}^{(1)} \mbox{\boldmath $\psi$}^{(0)}.
\end{equation}
Projecting the system (\ref{Lyapunovlinearsystem1}) onto the kernel
of ${\cal J} {\cal H}^{(0)}$, we find that the first-order
correction $\lambda_1$ is defined by the reduced eigenvalue problem:
\begin{equation}
\label{Lyapunovreducedeigenvalue1} 2 {\cal M}_1 {\bf c} =
\lambda_1^2 {\bf c},
\end{equation}
where ${\cal M}_1$ is given in (\ref{energyM1}). This result is in
agreement with the leading-order behavior (\ref{leading-order}) of
Lemma \ref{proposition-stability}. The matrix ${\cal M}_1$ 
has the same structure as in the
perturbation theory of continuous multi-pulse solitons \cite{S98}.
Therefore, the number of positive and negative eigenvalues of ${\cal
M}_1$ is defined by the following lemma.

\begin{Lemma}
\label{lemma-spliting} Let $n_0$, $z_0$, and $p_0$ be the numbers of
negative, zero and positive terms of $a_n = \cos(\theta_{n+1} -
\theta_n)$, $1 \leq n \leq N-1$, such that $n_0 + z_0 + p_0 = N-1$.
The matrix ${\cal M}_1$, defined by (\ref{Melements1}), has exactly
$n_0$ negative eigenvalues, $z_0+1$ zero eigenvalues, and $p_0$
positive eigenvalues.
\end{Lemma}

\begin{Proof}
See Lemma 5.4 and Appendix C of \cite{S98} for the proof.
\end{Proof}

When $z_0 = 0$, the zero eigenvalue of ${\cal M}_1$ with the
eigenvector $(1,1,...,1)^T$ is unique. Since all 
$\theta_n = \{0,\pi\}$, $n \in S$, then all $a_n \neq 0$, 
$1 \leq j \leq N-1$, such that $z_0 = 0$ and the splitting 
of the semi-simple zero eigenvalue of ${\cal H}^{(0)}$ is 
generic in the first-order of $\epsilon$ for the set $S_1$. By Lemma
\ref{lemma-spliting}, stability and instability of the discrete 
solitons in the set $S_1$
are defined in terms of the number $n_0$ of $\pi$-differences in
$\theta_{n+1} - \theta_n$ for $1 \leq n \leq N-1$. This result is in
agreement with Lemma \ref{proposition-n0} and Corollary
\ref{corollary-closure} for the family $S_1$. Thus, Theorem
\ref{theorem-stability} for the set $S_1$ 
is verified with explicit perturbation
series results.

We illustrate the stability results with two elementary examples of
the discrete solitons in the set $S_1$: $N = 2$ and $N = 3$. 
In the case $N = 2$, the discrete
two-pulse solitons consist of the Page mode (a) and the twisted mode
(b) as follows:
\begin{equation}
\mbox{\rm (a)} \; \theta_1 = \theta_2 = 0, \qquad \mbox{\rm (b)}
\; \theta_1 = 0, \; \theta_2 = \pi.
\end{equation}
The eigenvalues of matrix ${\cal M}_1$ are given explicitly as
$\gamma_1 = 0$ and $\gamma_2 = 2 \cos(\theta_2 - \theta_1)$.
Therefore, the Page mode (a) has one real unstable eigenvalue
$\lambda \approx 2 \sqrt{\epsilon}$ in the stability problem
(\ref{eigenvalue}) for small $\epsilon > 0$, while the twisted
mode (b) has no unstable eigenvalues but a simple pair of purely
imaginary eigenvalues $\lambda \approx \pm 2 i \sqrt{\epsilon}$ with
negative Krein signature. The latter pair may bifurcate to the
complex plane as a result of the
Hamiltonian Hopf bifurcation.

\begin{figure}[tbp]
\begin{center}
\epsfxsize=10.0cm %\centerline{}
\epsffile{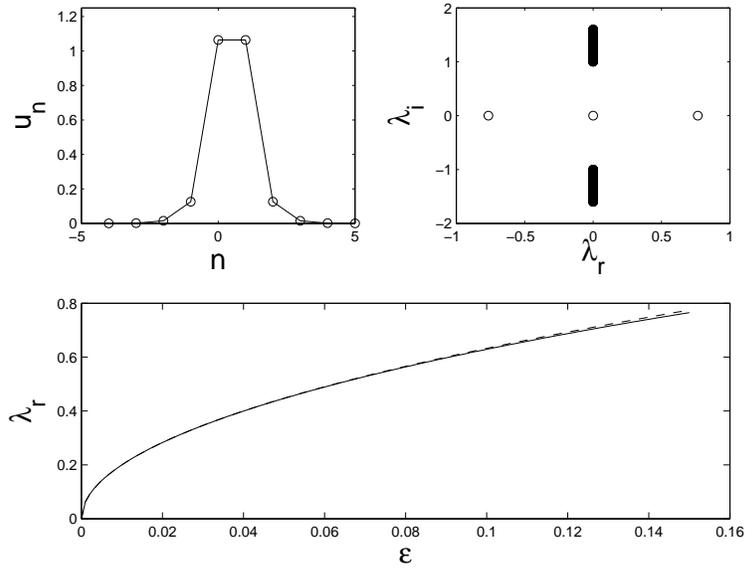}
%\end{tabular}
\caption{The top panel shows the spatial profile of
the Page mode (left) and the corresponding
spectral plane of the linear stability problem (right) 
for $\epsilon=0.15$.
The bottom subplot shows the continuation of the branch from
$\epsilon=0$ to $\epsilon=0.15$ and real positive eigenvalue
theoretically (dashed line) and numerically (solid line).} \label{depf1}
\end{center}
\end{figure}

\begin{figure}[tbp]
\begin{center}
\epsfxsize=10.0cm %\centerline{}
\epsffile{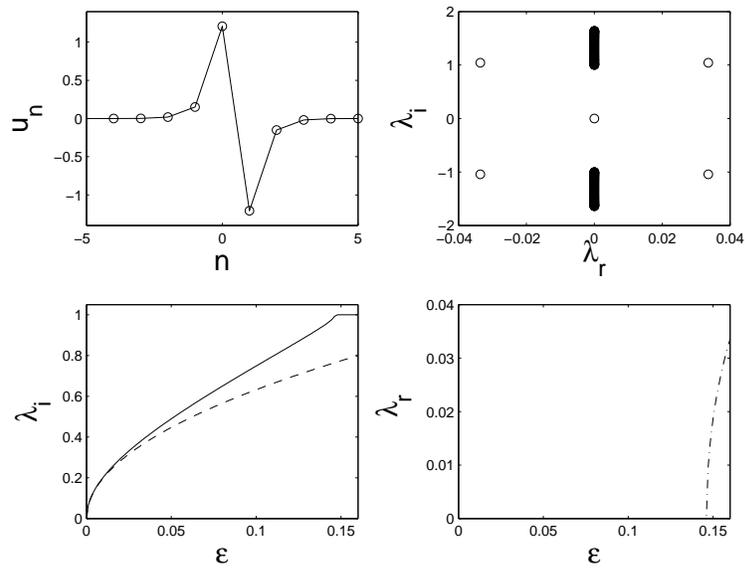}
%\end{tabular}
\caption{The top panel shows the twisted mode and the spectral plane
for $\epsilon = 0.15$. The bottom subplot shows the imaginary and
real parts of the eigenvalue with negative Krein signature, which
bifurcates to the complex plane at $\epsilon \approx 0.146$. }
\label{depf1a}
\end{center}
\end{figure}

These results are illustrated in Figures \ref{depf1} and
\ref{depf1a}, in agreement with numerical computations of the full
problems (\ref{1difference}) and (\ref{a-b-problem}). 
Fig. \ref{depf1} shows
the Page mode, while Fig. \ref{depf1a} corresponds to the twisted
mode. The top subplots of each figure show the mode profiles (left)
and the spectral plane $\lambda = \lambda_r + i \lambda_i$ of the
linear eigenvalue problem (right) for $\epsilon=0.15$. The bottom
subplots indicate the corresponding real (for the Page mode) and
imaginary (for the twisted mode) eigenvalues from the theory (dashed
line) versus the full numerical result (solid line). We find the
agreement between the theory and the numerical computation to be
excellent in the case of the Page mode (Fig. \ref{depf1}). For the
twisted mode (Fig. \ref{depf1a}), the agreement is within
the $5\%$-error for $\epsilon<0.0258$. For larger values of
$\epsilon$, the difference between the theory and numerics grows. 
The imaginary eigenvalues collide at $\epsilon \approx 0.146$ 
with the band edge of the continuous spectrum, such that 
the real part $\lambda_r$ becomes non-zero for $\epsilon > 0.146$.

\begin{figure}[tbp]
\begin{center}
\epsfxsize=7.0cm %\centerline{}
\epsffile{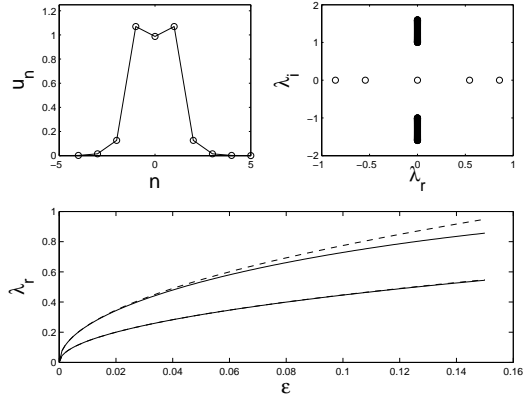}
%\end{tabular}
\caption{Same as Fig. \ref{depf1} for the mode (a) 
with three excited sites in phase.} \label{depf2}
\end{center}
\end{figure}

\begin{figure}[tbp]
\begin{center}
\epsfxsize=7.0cm %\centerline{}
\epsffile{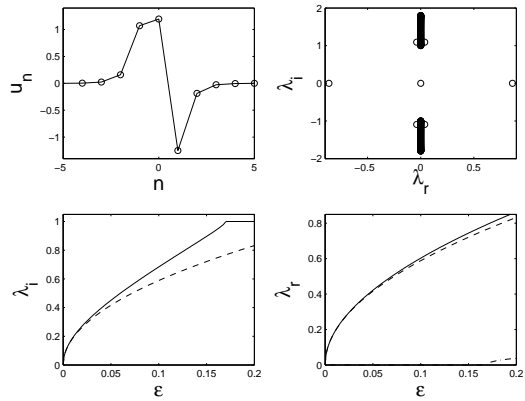}
%\end{tabular}
\caption{Same as Fig. \ref{depf1a} for
the mode (b) with three excited
sites, where the left and middle sites are in phase and the right
$\pi$ is out of phase.}
\label{depf2a}
\end{center}
\end{figure}

\begin{figure}[tbp]
\begin{center}
\epsfxsize=7.0cm %\centerline{}
\epsffile{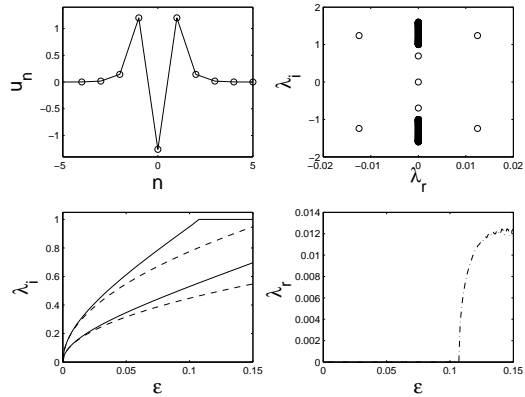}
%\end{tabular}
\caption{Same as Fig. \ref{depf1a} for
the mode (c) with three excited
sites, where adjacent sites are out of phase with each other.} 
\label{depf2b}
\end{center}
\end{figure}

In the case $N = 3$, the discrete three-pulse solitons consist of
the three modes as follows:
\begin{equation}
\label{three-cases} \mbox{\rm (a)} \; \theta_1 = \theta_2 = \theta_3
= 0, \qquad \mbox{\rm (b)} \; \theta_1 = \theta_2 = 0, \; \theta_3 =
\pi, \qquad \mbox{\rm (c)} \; \theta_1 = 0, \; \theta_2 = \pi, \;
\theta_3 = 0.
\end{equation}
The eigenvalues of matrix ${\cal M}_1$ are given explicitly as
$\gamma_1 = 0$ and
$$
\gamma_{2,3} = \cos(\theta_2 - \theta_1) + \cos(\theta_3 - \theta_2)
\pm \sqrt{\cos^2(\theta_2 - \theta_1) - \cos(\theta_2 - \theta_1)
\cos(\theta_3 - \theta_2) + \cos^2(\theta_3-\theta_2)}.
$$
The mode (a) has two real unstable eigenvalues $\lambda \approx
\sqrt{6 \epsilon}$ and $\lambda \approx \sqrt{2 \epsilon}$ in the
stability problem (\ref{eigenvalue}) for small $\epsilon > 0$.
The mode (b) has one real unstable eigenvalue $\lambda \approx
\sqrt{2 \sqrt{3} \epsilon}$ and a simple pair of purely imaginary
eigenvalues $\lambda \approx \pm i \sqrt{2 \sqrt{3} \epsilon}$ with
negative Krein signature. This pair may bifurcate to the complex
plane as a result of the Hamiltonian Hopf
bifurcation. The mode (c) has no unstable eigenvalues but two pairs
of purely imaginary eigenvalues $\lambda \approx \pm i \sqrt{6
\epsilon}$ and $\lambda \approx \pm i \sqrt{2 \epsilon}$ with
negative Krein signature. The two pairs may bifurcate to the complex
plane as a result of the two successive Hamiltonian Hopf bifurcations.

Figures \ref{depf2}--\ref{depf2b} summarize the results for the
three modes (a)--(c), given in (\ref{three-cases}). 
Fig. \ref{depf2} corresponds to the
mode (a), where two real positive
eigenvalues give rise to instability for any $\epsilon \neq 0$. 
The error between theoretical and numerical results is within 
$5\%$ for $\epsilon < 0.15$ for one real eigenvalue and 
for $\epsilon < 0.0865$ for the other eigenvalue. 
Similar results are observed on Fig. \ref{depf2a} for the mode (b),
where the real positive eigenvalue and a pair of imaginary eigenvalues
with negative Krein signature are generated for $\epsilon > 0$.
The imaginary eigenvalue collides with the band edge of the
continuous spectrum at $\epsilon \approx 0.169$, which results in
the Hamiltonian Hopf bifurcation. Finally, Fig. \ref{depf2b} shows
the mode (c), where
two pairs of imaginary eigenvalues with negative Krein signature
exist for $\epsilon > 0$. The first Hamiltonian Hopf bifurcation
occurs for $\epsilon \approx 0.108$, while the second one occurs for
much larger values of $\epsilon \approx 0.223$, which is beyond the
scale of Fig. \ref{depf2b}.

\section{Bifurcations of the discrete solitons in the set $S_2$}

Here we study the set $S_2$ with the revised perturbation series 
expansions. The solution is defined by the power
series (\ref{perturbation1}), where the zero-order term
$\phi_n^{(0)}$ is given by (\ref{soliton}) with $\theta_n =
\{0,\pi\}$ for all $n \in S$ and the first-order term $\phi_n^{(1)}$
is given by
\begin{eqnarray}
\nonumber & \phantom{=} & \phi_n^{(1)} = e^{i \theta_{n+1}} + e^{i
\theta_{n-1}}, \quad n = 2m, \; 1 \leq m \leq N-1, \\
\label{solution2} & \phantom{=} & \phi_0^{(1)} = e^{i \theta_1},
\qquad \phi_{2N}^{(1)} = e^{i \theta_{2N-1}},
\end{eqnarray}
while all other elements of $\phi_n^{(1)}$ are empty. The
second-order term $\phi_n^{(2)}$ solves the inhomogeneous problem:
\begin{equation}
\label{system2} (1 - 3 \phi_n^{(0) 2} ) \phi_n^{(2)} =
\phi_{n+1}^{(1)} + \phi_{n-1}^{(1)} + 3 \phi_n^{(1) 2} \phi_n^{(0)},
\end{equation}
with the unique solution:
\begin{eqnarray}
\nonumber & \phantom{=} & \phi_n^{(2)} = - \frac{1}{2} \left(
\cos(\theta_{n+2}-\theta_n) + \cos(\theta_{n-2} - \theta_n) + 2
\right) e^{i \theta_n}, \quad n = 2m-1, \; 2 \leq m \leq N-1, \\
\nonumber & \phantom{=} & \phi_1^{(2)} = - \frac{1}{2} \left(
\cos(\theta_{3}-\theta_1) + 2 \right) e^{i \theta_1}, \qquad
\phi_{2N-1}^{(2)} = - \frac{1}{2} \left(
\cos(\theta_{2N-1}-\theta_{2N-3}) + 2 \right) e^{i \theta_{2N-1}}, \\
\label{solution20} & \phantom{=} & \phi_{-1}^{(2)} = e^{i\theta_1},
\qquad \phi_{2N+1}^{(2)} = e^{i \theta_{2N-1}},
\end{eqnarray}
while all other elements of $\phi_n^{(2)}$ are empty. The symmetric
matrix ${\cal H}$ is defined by the power series (\ref{matrix1}),
where the zero-order term ${\cal H}^{(0)}$ is given by
(\ref{energy-0}) and the first-order term ${\cal H}^{(1)}$ is given
by (\ref{energy1}), where $\phi_n^{(0)} \phi_n^{(1)} = 0$, $n \in
\Z$. The second-order term ${\cal H}^{(2)}$ has the structure:
\begin{equation}
\nonumber {\cal H}_{n,n}^{(2)} = -2 \phi_n^{(0)} \phi_n^{(2)} \left(
\begin{array}{cc} 3 & 0 \\ 0 & 1 \end{array} \right), \quad
n = 2m -1, \; 1 \leq m \leq N
\end{equation}
and
\begin{equation}
\nonumber {\cal H}_{n,n}^{(2)} = - \phi_n^{(1) 2} \left(
\begin{array}{cc} 3 & 0 \\ 0 & 1 \end{array} \right),
\qquad n = 2m, \; 0 \leq m \leq N,
\end{equation}
while all other blocks of ${\cal H}_{n,m}^{(2)}$ are empty.
Similarly to the previous section, the semi-simple zero eigenvalue
of the problem ${\cal H} \mbox{\boldmath $\varphi$} = \gamma
\mbox{\boldmath $\varphi$}$ is split as $\epsilon > 0$ according
to the modified perturbation series expansion:
\begin{eqnarray}
\label{eigenvector2} \mbox{\boldmath $\varphi$} = \mbox{\boldmath
$\varphi$}^{(0)} + \epsilon \mbox{\boldmath $\varphi$}^{(1)} +
\epsilon \mbox{\boldmath $\varphi$}^{(2)} + {\rm O}(\epsilon^3),
\qquad \gamma = \epsilon^2 \gamma_2 + {\rm O}(\epsilon^3),
\end{eqnarray}
where the zero-order term $\mbox{\boldmath $\varphi$}^{(0)}$ is
given by (\ref{zeroeigenvector1}) and the first-order term $\mbox{\boldmath
$\varphi$}^{(1)}$ has the form:
\begin{equation}
\mbox{\boldmath $\varphi$}^{(1)} = \sum_{n \in S} c_n \left( {\cal
S}_+ {\bf f}_n + {\cal S}_- {\bf f}_n \right),
\end{equation}
where ${\cal S}_{\pm}$ are shift operators of the non-zero $2$-block
of ${\bf f}_n$ up and down. The second-order term $\mbox{\boldmath
$\varphi$}^{(2)}$ is found from the inhomogeneous system:
\begin{equation}
\label{linearsystem2} {\cal H}^{(0)} \mbox{\boldmath
$\varphi$}^{(2)} = \gamma_2 \mbox{\boldmath $\varphi$}^{(0)} -
{\cal H}^{(1)} \mbox{\boldmath $\varphi$}^{(1)} - {\cal H}^{(2)}
\mbox{\boldmath $\varphi$}^{(0)}.
\end{equation}
Projecting the system (\ref{linearsystem2}) onto the kernel of
${\cal H}^{(0)}$, we find the reduced eigenvalue problem:
\begin{equation}
\label{reducedeigenvalue2} {\cal M}_2 {\bf c} = \gamma_2 {\bf c},
\end{equation}
where ${\bf c} = (c_1,c_3,...,c_{2N-1})^T$ and ${\cal M}_2$ is the
tri-diagonal $N$-by-$N$ matrix, given by
\begin{equation}
\label{energyM2} \left( {\cal M}_2 \right)_{m,n} = \left( {\bf
f}_{2m-1}, {\cal H}^{(2)} {\bf f}_{2n-1} \right) + \left( {\bf
f}_{2m-1}, {\cal H}^{(1)} ({\cal S}_+ + {\cal S}_-) 
{\bf f}_{2n-1} \right),
\end{equation}
for $1 \leq n,m \leq N$, or explicitly, based on the first-order and
second-order solutions (\ref{solution2}) and (\ref{solution20}):
\begin{eqnarray}
\nonumber (M_2)_{n,n} = \cos(\theta_{2n+1} -
\theta_{2n-1}) + \cos(\theta_{2n-3} - \theta_{2n-1}), \qquad 1 < n < N, \\
\nonumber
(M_2)_{n,n+1} = (M_2)_{n+1,n} = - \cos(\theta_{2n+1}-
\theta_{2n-1}), \qquad 1 \leq n < N, \\
\label{Melements2}
(M_2)_{1,1} = \cos(\theta_3 - \theta_1), \qquad (M_2)_{N,N} =
\cos(\theta_{2N-1} - \theta_{2N-3}).
\end{eqnarray}
Similarly, the multiple zero eigenvalue of the problem ${\cal J H}
\mbox{\boldmath $\psi$} = \lambda \mbox{\boldmath $\psi$}$ is split
as $\epsilon > 0$ according to the modified perturbation series
expansion:
\begin{eqnarray}
\label{Lyapunoveigenvector2} \mbox{\boldmath $\psi$} =
\mbox{\boldmath $\psi$}^{(0)} + \epsilon \mbox{\boldmath
$\psi$}^{(1)} + \epsilon^2 \mbox{\boldmath $\psi$}^{(2)} + {\rm
O}(\epsilon^3), \qquad \lambda = \epsilon \lambda_1 + \epsilon^2
\lambda_2 + {\rm O}(\epsilon^3),
\end{eqnarray}
where the zero-order term $\mbox{\boldmath $\psi$}^{(0)} =
\mbox{\boldmath $\varphi$}^{(0)}$ is given by
(\ref{zeroeigenvector1}) and the first-order term $\mbox{\boldmath
$\psi$}^{(1)}$ has the form:
\begin{equation}
\label{Lyapunovzeroeigenvector2} \mbox{\boldmath $\psi$}^{(1)} =
\sum_{n \in S} c_n \left( {\cal S}_+ {\bf f}_n + {\cal S}_- {\bf
f}_n \right) + \frac{\lambda_1}{2} \sum_{n \in S} c_n {\bf g}_n.
\end{equation}
The second-order term $\mbox{\boldmath $\psi$}^{(2)}$ is found
from the inhomogeneous system:
\begin{equation}
\label{Lyapunovlinearsystem2} {\cal J} {\cal H}^{(0)}
\mbox{\boldmath $\psi$}^{(2)} = \lambda_1 \mbox{\boldmath
$\psi$}^{(1)} + \lambda_2 \mbox{\boldmath $\psi$}^{(0)} - {\cal J}
{\cal H}^{(1)} \mbox{\boldmath $\psi$}^{(1)} - {\cal J} {\cal
H}^{(2)} \mbox{\boldmath $\psi$}^{(0)}.
\end{equation}
Projecting the system (\ref{Lyapunovlinearsystem2}) onto the kernel
of ${\cal J} {\cal H}^{(0)}$, we find the reduced eigenvalue
problem:
\begin{equation}
\label{Lyapunovreducedeigenvalue2} 2 {\cal M}_2 {\bf c} =
\lambda_1^2 {\bf c},
\end{equation}
in accordance with Lemma \ref{proposition-stability}. Since the
matrix ${\cal M}_2$ has exactly the structure of the matrix ${\cal
M}_1$, described in Lemma \ref{lemma-spliting}, we conclude that the
stability and instability of the discrete solitons in the set 
$S_2$ is defined in terms of the number $n_0$ of
$\pi$-differences in $\theta_{2n+1} - \theta_{2n-1}$, $1 \leq n \leq
N-1$, in accordance with Lemma \ref{proposition-n0} and Corollary
\ref{corollary-closure}. Thus, Theorem \ref{theorem-stability} 
for the set $S_2$ is verified with explicit perturbation series results.

\begin{figure}[tbp]
\begin{center}
\epsfxsize=10.0cm %\centerline{}
\epsffile{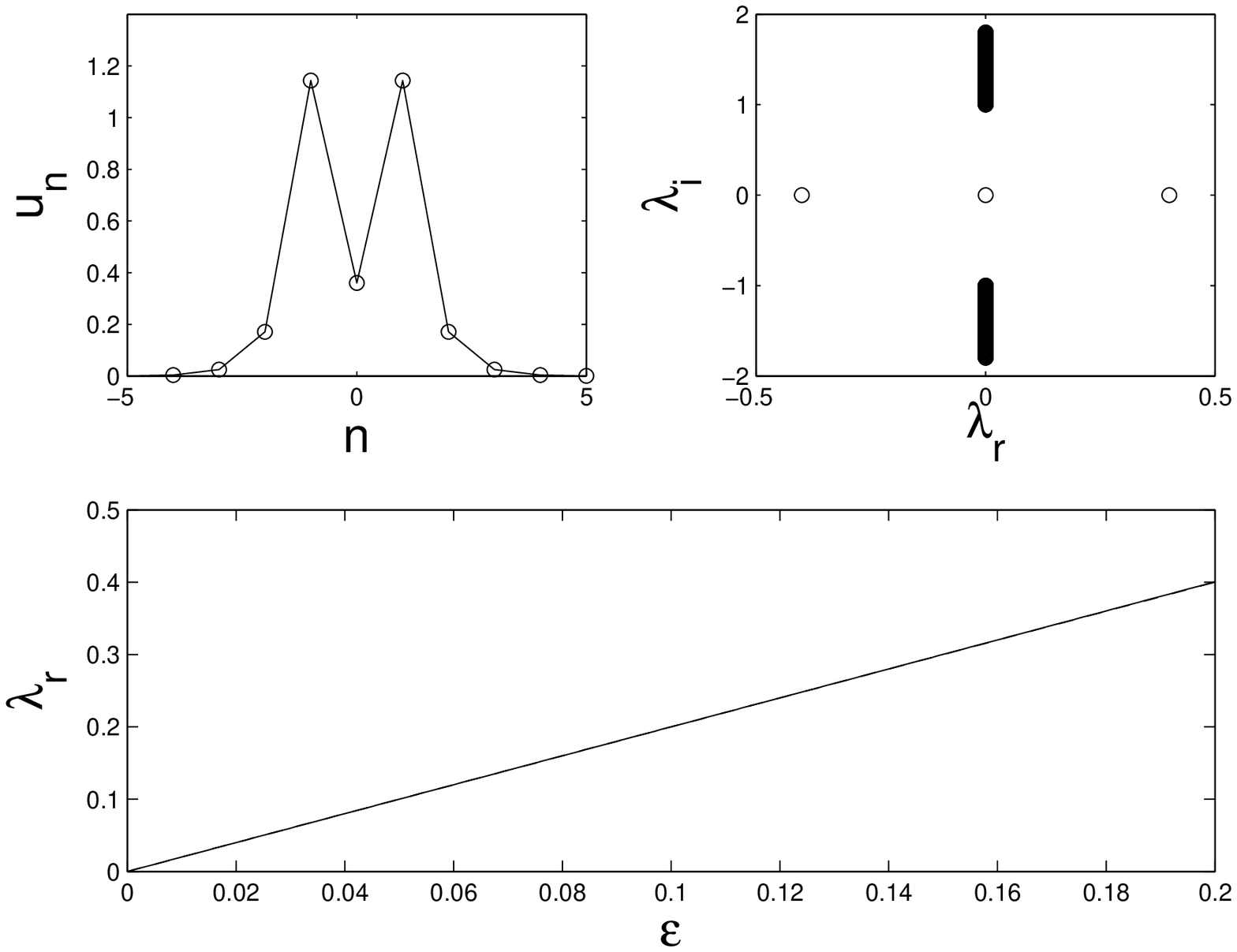}
%\end{tabular}
\caption{Same as Fig. \ref{depf1} but for the Page mode of the
set $S_2$.} \label{depf3}
\end{center}
\end{figure}

\begin{figure}[tbp]
\begin{center}
\epsfxsize=10.0cm %\centerline{}
\epsffile{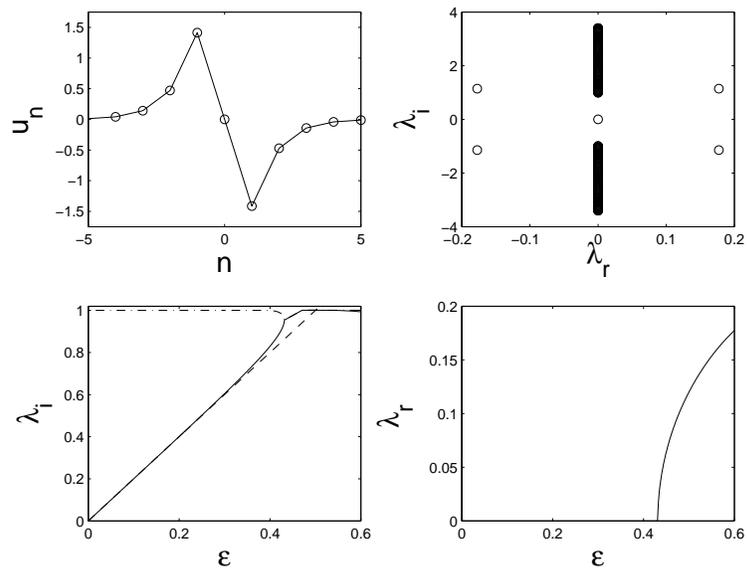}
%\end{tabular}
\caption{Same as Fig. \ref{depf1a} but for the twisted mode of the
set $S_2$.} \label{depf3a}
\end{center}
\end{figure}

\begin{figure}[tbp]
\begin{center}
\epsfxsize=7.0cm %\centerline{}
\epsffile{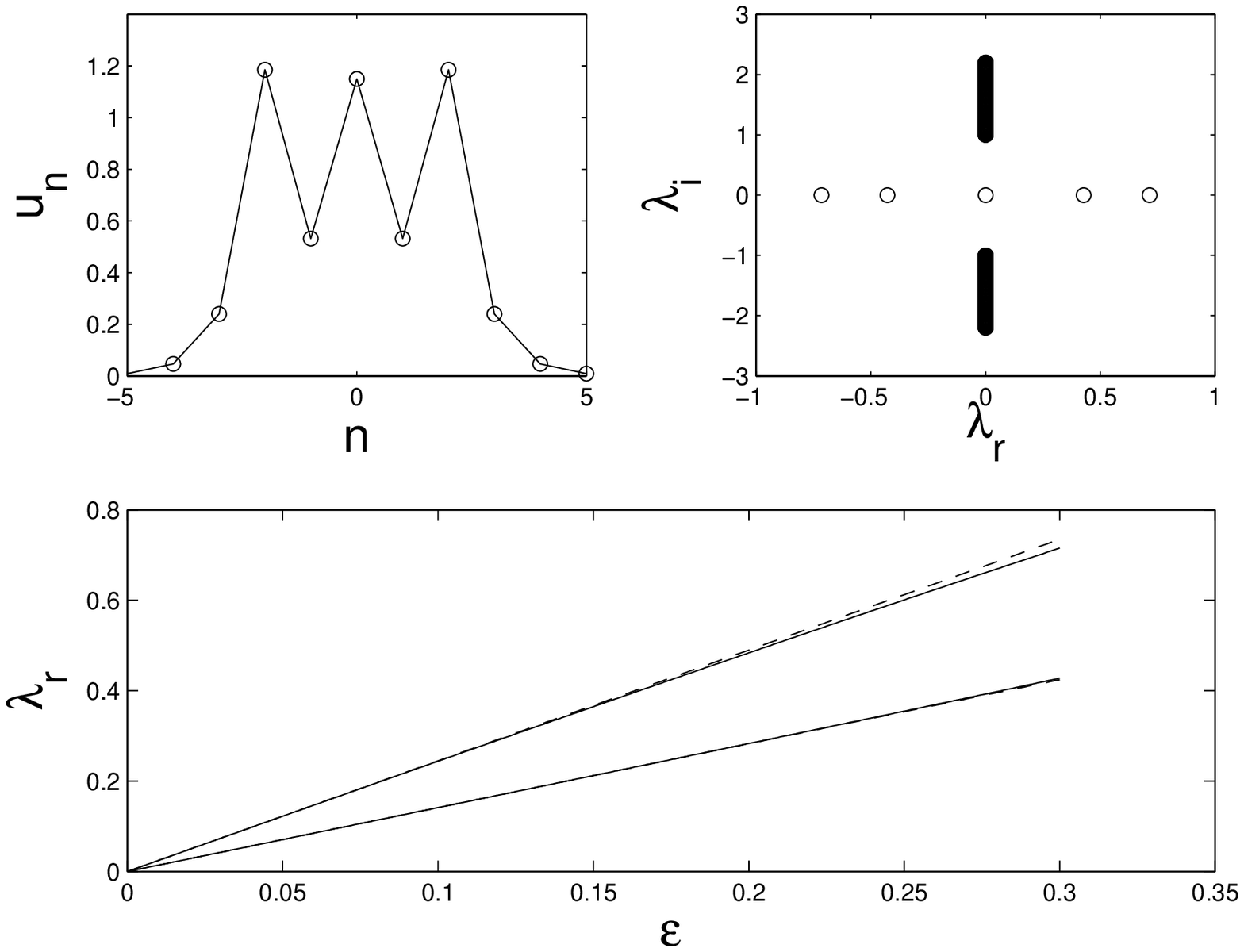}
%\end{tabular}
\caption{Same as Fig. \ref{depf2} but for the mode (a) of the set
$S_2$.} \label{depf4}
\end{center}
\end{figure}

\begin{figure}[tbp]
\begin{center}
\epsfxsize=7.0cm %\centerline{}
\epsffile{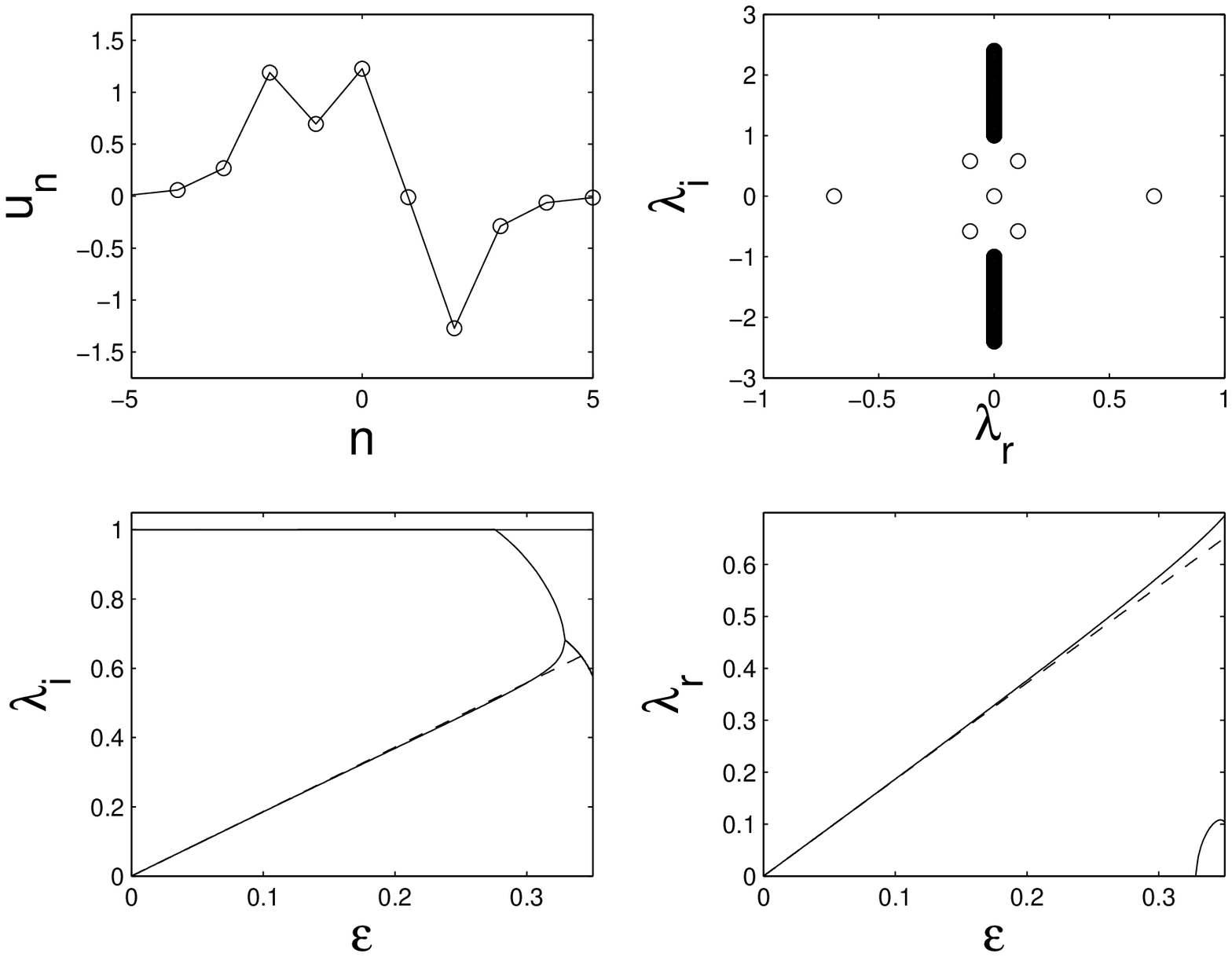}
%\end{tabular}
\caption{Same as Fig. \ref{depf2a} but for the mode (b) of the
set $S_2$.} \label{depf4a}
\end{center}
\end{figure}

\begin{figure}[tbp]
\begin{center}
\epsfxsize=7.0cm %\centerline{}
\epsffile{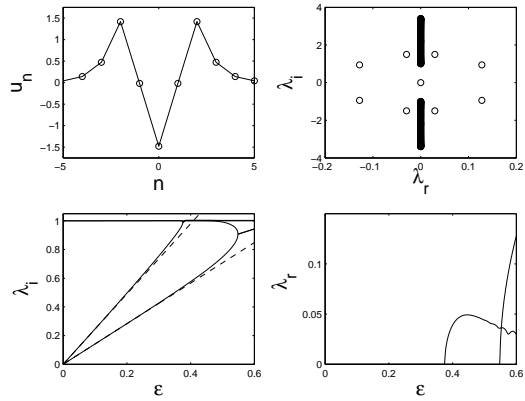}
%\end{tabular}
\caption{Same as Fig. \ref{depf2b} but for the mode (c) of the
set $S_2$.} \label{depf4b}
\end{center}
\end{figure}

We summarize that bifurcations and stability of the 
discrete solitons in the set $S_2$ is exactly equivalent to those
in the set $S_1$, but the splitting of all zero eigenvalues
occurs in the order of $\epsilon^2$, rather than in the order of
$\epsilon$. These results for the set $S_2$ with $N = 2$ and $N =
3$ are shown on Figures \ref{depf3}--\ref{depf4b}, in full analogy
with those for the set $S_1$. The corresponding asymptotic
approximations of eigenvalues can be ``translated'' from those of
the previous section by substituting $\sqrt{\epsilon} \rightarrow
\epsilon$. Fig. \ref{depf3} shows the Page mode where the agreement
with the theory is excellent for $\epsilon < 0.2$. Fig. \ref{depf3a}
shows the twisted mode with very good agreement for $\epsilon <
0.415$ and the Hamiltonian Hopf bifurcation at $\epsilon \approx
0.431$. The only difference from the twisted mode of Fig.
\ref{depf1a} is that the imaginary eigenvalue of negative Krein
signature collides with the imaginary eigenvalue of positive Krein
signature, rather than with the band edge of the continuous
spectrum. Figs. \ref{depf4}, \ref{depf4a}, and \ref{depf4b} show the
modes (a), (b), and (c), respectively, of the three excited sites.
Again, the Hamiltonian Hopf bifurcations occur when the imaginary
eigenvalues of negative Krein signature collide with the imaginary
eigenvalues of positive Krein signature. For the mode (b), the
bifurcation occurs at $\epsilon \approx 0.328$ (see Fig.
\ref{depf4a}). For the mode (c), two bifurcations occur at $\epsilon
\approx 0.375$ and $\epsilon \approx 0.548$ (see Fig. \ref{depf4b}).

\section{Summary}

We have studied stability of discrete solitons in the one-dimensional
NLS lattice (\ref{dNLS}) with $d = 1$. We have rigorously proved the
numerical conjecture that the discrete solitons with anti-phase
excited nodes are stable near the anti-continuum limit, while all
other discrete solitons are linearly unstable with real positive
eigenvalues in the stability problem. 
Additionally, we gave a precise count of the real eigenvalues and 
pairs of imaginary eigenvalues with negative Krein signature.
These results are not affected
if the excited nodes are separated by an arbitrary sequence of empty
nodes. We studied two particular sets of discrete solitons with
explicit perturbation series expansions and numerical approximations and found
very good agreement between the asymptotic and numerical computations.

Stability and instability results remain invariant if the discrete
solitons are excited in the two-dimensional NLS lattice (\ref{dNLS})
with $d = 2$, such that the set $S$ is an open discrete 
contour on the plane. Similar perturbation
series expansions for the sets $S_1$ and $S_2$ in the
two-dimensional NLS lattice can be developed and the same matrices
${\cal M}_1$ and ${\cal M}_2$ define stability and instability of
these discrete solitons.

In the second forthcoming paper of this series, we shall consider
a closed discrete contour on the plane for the set $S$. Such sets 
may support both discrete solitons and 
discrete vortices with a non-zero topological charge. Continuation
of the limiting solutions from $\epsilon = 0$ to $\epsilon \neq 0$
is a non-trivial problem if the amplitudes $\phi_n$ are complex-valued. 
We shall study persistence,
multiplicity and stability of such continuations with the methods of
Lyapunov--Schmidt reductions.

\vspace{5mm}

{\bf Acknowledgements} This work was partially supported by 
NSF-DMS-0204585, NSF-CAREER, and the Eppley Foundation for Research (PGK).
PGK is also particularly grateful to G.L. Alfimov, V.A. Brazhnyi, and 
V.V. Konotop for numerous insightful discussions regarding discrete 
soliton stability and for their communication of details 
of their work \cite{ABK04} prior to publication.


\begin{thebibliography}{99}

\bibitem{EMSAPA02} H.S. Eisenberg, R. Morandotti, Y. Silberberg, J.M. Arnold, G. Pennelli, and J.S. Aitchison,
Optical discrete solitons in waveguide arrays. I. Soliton formation,
J. Opt. Soc. Am. B {\bf 19} (2002) 2938-2944.

\bibitem{PMAAESPL02} U. Peschel, R. Morandotti, J.M. Arnold, J.S. Aitchison, H.S. Eisenberg, Y. Silberberg, T. Pertsch, and F. Lederer,
Optical discrete solitons in waveguide arrays. 2. Dynamic
properties, J. Opt. Soc. Am. B {\bf 19} (2002) 2637-2644.

\bibitem{ESCFS02} N.K. Efremidis, S. Sears, D.N. Christodoulides, J.W. Fleischer, and M. Segev, Discrete solitons
in photorefractive optically induced photonic lattices, Phys. Rev. E
{\bf 66} (2002) 046602.

\bibitem{SKEA03} A.A. Sukhorukov, Yu.S. Kivshar, H.S. Eisenberg, and Y. Silberberg,
Spatial optical solitons in waveguide arrays, IEEE J. Quantum Elect. {\bf 39} (2003) 31-50.

\bibitem{CBFMMTSI01} F.S.\ Cataliotti, S. Burger, C. Fort, P. Maddaloni, F. Minardi,
A. Trombettoni, A. Smerzi, and M. Inguscio,
Josephson junction arrays with Bose-Einstein condensates,
\newblock Science {\bf  293} (2001) 843-846.

\bibitem{CFFFMI03} F.S. Cataliotti, L. Fallani, F. Ferlaino, C. Fort, P. Maddaloni, and
M. Inguscio, Superfluid current disruption in a chain of weakly
coupled Bose-Einstein condensates, New. J. Phys. {\bf 5} (2003) 71.

\bibitem{ABDKS01} F.Kh.\ Abdullaev, B.B. Baizakov, S.A. Darmanyan, V.V. Konotop, and M. Salerno,
Nonlinear excitations in arrays of Bose-Einstein condensates, Phys.\
Rev.\ A {\bf 64} (2001) 043606.

\bibitem{AKKS02} G.L. Alfimov, P.G. Kevrekidis, V.V. Konotop, and M. Salerno,
Wannier functions analysis of the nonlinear Schrodinger equation
with a periodic potential, Phys. Rev. E {\bf 66} (2002) 046608.

\bibitem{F03} M.V. Fistul, Resonant breather states in Josephson
coupled systems, Chaos {\bf 13}, 725--732 (2003)

\bibitem{MO03} J.J. Mazo and T.P. Orlando, Discrete breathers in
Josephson arrays, Chaos {\bf 13}, 733-743 (2003)

%\bibitem{PB89} M. Peyrard and A.R. Bishop, Statistical-mechanics of a nonlinear model for DNA denaturation, Phys. Rev. Lett., {\bf 62} (1989) 2755-2758.

\bibitem{DPB93} T. Dauxois, M. Peyrard and A.R. Bishop, Entropy-driven DNA denaturation,
Phys. Rev. E {\bf 47} (1993) R44-R47.

\bibitem{PDHW93} M. Peyrard, T. Dauxois, H. Hoyet, and C.R. Willis,
Biomolecular dynamics of DNA--Statistical-mechanics and dynamical models,
\newblock Physica D {\bf 68} (1993) 104-115.

\bibitem{A97} S. Aubry, Breathers in nonlinear lattices: existence, linear stability and quantization, Physica D {\bf 103} (1997) 201-250.

\bibitem{FW98} S. Flach and C.R. Willis, Discrete breathers, Physics Reports {\bf 295} (1998) 181-264.

\bibitem{HT99} D. Hennig and G. Tsironis, Wave transmission in nonlinear lattices, Physics Reports {\bf 307} (1999) 333-432.

\bibitem{KRB01} P.G. Kevrekidis, K.O. Rasmussen, and A.R. Bishop,
The discrete nonlinear Schrodinger equation: A survey of recent
results, Int. J. Mod. Phys. B {\bf 15} (2001) 2833-2900.

\bibitem{EJ03} J.Ch. Eilbeck and M. Johansson, in {\em Localization and Energy Transfer in Nonlinear Systems}, L. Vazquez, R.S. MacKay, and M.P. Zorzano (eds.), (World Scientific,
Singapore, 2003), p.44.

%\bibitem{KF04} P.G. Kevrekidis and D.J. Frantzeskakis,
%Pattern forming dynamical instabilities of Bose-Einstein condensates,
%Mod. Phys. Lett. B {\bf 18} (2004) 173-202.

\bibitem{PK92} M. Peyrard and Yu.S. Kivshar, Modulational Instabilities in discrete
lattices, Phys. Rev. A {\bf 46} (1992) 3198-3205.


\bibitem{ABK04} G.L. Alfimov, V.A. Brazhnyi, and V.V. Konotop,
On classification of intrinsic localized modes for the discrete nonlinear
Schr\"{o}dinger equation, Physica D {\bf 194} (2004) 127-150.


\bibitem{DM00} H.R. Dullin and J.D. Meiss, Generalized Henon maps: the cubic diffeomorphisms
of the plane, Physica D {\bf 143} (2000) 262-289.

\bibitem{BBJ00} J.M. Bergamin, T. Bountis, and C. Jung, A method for locating symmetric homoclinic
orbits using symbolic dynamics, J. Phys. A: Math. Gen. {\bf 33} (2000) 8059-8070.

\bibitem{BBV02} J.M. Bergamin, T. Bountis, and M.N. Vrahatis,
Homoclinic orbits of invertible maps, Nonlinearity {\bf 15} (2002) 1603-1619.

\bibitem{AA90} S. Aubry and G. Abramovici, Chaotic trajectories in the standard map-the concept
of antiintegrability, Physica D {\bf 43} (1990) 199-219.

\bibitem{MA94} R.S. MacKay and S. Aubry, Proof of existence of breathers for time-reversible
or Hamiltonian networks of weakly coupled oscillators, Nonlinearity {\bf 7} (1994) 1623-1643.

\bibitem{N74} L. Nirenberg, {\em Topics in Nonlinear Functional Analysis} (Courant Institute, NY, 1974).

\bibitem{W99} M. Weinstein, Excitation thresholds for nonlinear localized
modes on lattices, Nonlinearity {\bf 12} (1999) 673-691.


\bibitem{KK01} T. Kapitula and P.G. Kevrekidis, Stability of waves in discrete systems,
Nonlinearity {\bf 14} (2001) 533-566.

\bibitem{JA97} M. Johansson and S. Aubry, Existence and stability of quasiperiodic breathers
in the discrete nonlinear Schr\"{o}dinger equation, Nonlinearity {\bf 10} (1997) 1151-1178.

\bibitem{KBR01} P.G. Kevrekidis, A.R. Bishop, and K.O. Rasmussen, Twisted localized modes,
Phys. Rev. E {\bf 63} (2001) 036603.

\bibitem{KKM01} T. Kapitula, P.G. Kevrekidis, and B.A. Malomed, Stability of multiple pulses
in discrete systems, Phys. Rev. E {\bf 63} (2001) 036604.

\bibitem{KW03} P.G. Kevrekidis and M.I. Weinstein, Breathers on a background: periodic and
quasiperiodic solutions of extended discrete nonlinear wave systems, Math. Comp. Simulat.
{\bf 62} (2003) 65-78.


\bibitem{MJKA02} A.M. Morgante, M. Johansson, G. Kopidakis, and S. Aubry,
Standing wave instabilities in a chain of nonlinear coupled oscillators, Physica D {\bf 162} (2002) 53-94.

\bibitem{J04} M. Johansson, Hamiltonian Hopf bifurcations in the discrete nonlinear
Schr\"{o}dinger equation, J. Phys. A: Math. Gen. {\bf 37} (2004) 2201-2222.

\bibitem{SJA97} B. Sandstede, C.K.R.T. Jones, and J.C. Alexander,
Existence and stability of $N$-pulses on optical fibers with
phase-sensitive amplifiers, Physica D {\bf 106}, 167--206 (1997)

\bibitem{S98} B. Sandstede, Stability of multiple-pulse solutions, Trans. Amer. Math. Soc. {\bf 350} (1998) 429-472.

\bibitem{K01} T. Kapitula, Stability of waves in perturbed Hamiltonian systems, Physica D
{\bf 156} (2001) 186-200.

\bibitem{KK04} T. Kapitula and P.G. Kevrekidis,
Linear stability of perturbed Hamiltonian systems: theory and a case example,
J. Phys. A: Math. Gen. {\bf 37} (2004) 7509-7526.

\bibitem{HJ85} R. Horn and C. Johnson, {\em Matrix Analysis}, (Cambridge University Press, 1985).

\bibitem{LL92} H. Levy and F. Lessman, {\em Finite Difference Equations} (Dover, New York, 1992).

\bibitem{P04} D. Pelinovsky, Inertia law for spectral stability of
solitary waves in coupled nonlinear Schr\"{o}dinger equations,
Proc. Roy. Soc. Lond. A, in press
(preprint available at
http://dmpeli.math.mcmaster.ca/PaperBank/diagonalization.pdf).


\bibitem{KKS04} T. Kapitula, P.G. Kevrekidis, and B. Sandstede,
Counting eigenvalues via the Krein signature in infinite-dimensional
Hamiltonian systems, Physica D {\bf 195} (2004) 263--282.





\end{thebibliography}
\end{document}